\documentclass[aps,prb,twocolumn,floatfix,superscriptaddress,showpacs,a4paper,citeautoscript]{revtex4}

\usepackage{graphicx}
\usepackage{bbm}
\usepackage{amsmath,amssymb}
\usepackage{graphicx}

\newcommand{\be}{\begin{equation}}
\newcommand{\beq}{\begin{equation}}
\newcommand{\ee}{\end{equation}}
\newcommand{\bea}{\begin{eqnarray}}
\newcommand{\eea}{\end{eqnarray}}
\newcommand{\ba}{\begin{array}}
\newcommand{\ea}{\end{array}}
\newcommand{\Id}[1] {\int \! \! {\rm d}^2 #1\:}

\renewcommand{\vr} {{\bf r}}

\newcommand{\vs} {{\bf s}}

\newcommand{\vj} {{\bf j}}

\begin{document}
\title{Density gradients for the exchange energy of electrons in two dimensions}

\author{Stefano Pittalis}
\email[Electronic address:\;]{pittalis@physik.fu-berlin.de}
\affiliation{Institut f{\"u}r Theoretische Physik,
Freie Universit{\"a}t Berlin, Arnimallee 14, D-14195 Berlin, Germany}
\affiliation{European Theoretical Spectroscopy Facility (ETSF)}

\author{Esa R{\"a}s{\"a}nen}
\email[Electronic address:\;]{erasanen@jyu.fi}
\affiliation{Institut f{\"u}r Theoretische Physik,
Freie Universit{\"a}t Berlin, Arnimallee 14, D-14195 Berlin, Germany}
\affiliation{European Theoretical Spectroscopy Facility (ETSF)}
\affiliation{Nanoscience Center, Department of Physics, 
University of Jyv{\"a}skyl{\"a}, FI-40014 Jyv{\"a}skyl{\"a}, Finland}

\author{Jos\'e G. Vilhena}
\affiliation{Laboratoire de Physique de la Mati\`{e}re
Condens\'{e} et Nanostructures,
Universit\'{e} Lyon I, CNRS, UMR 5586, Domaine scientifique de la
Doua, F-69622 Villeurbanne Cedex, France}
\affiliation{European Theoretical Spectroscopy Facility (ETSF)}

\author{Miguel A.\,L. Marques}
\email[Electronic address:\;]{marques@tddft.org}
\affiliation{Laboratoire de Physique de la Mati\`{e}re
Condens\'{e} et Nanostructures,
Universit\'{e} Lyon I, CNRS, UMR 5586, Domaine scientifique de la
Doua, F-69622 Villeurbanne Cedex, France}
\affiliation{European Theoretical Spectroscopy Facility (ETSF)}

\date{\today}

\begin{abstract}
We derive a generalized gradient approximation to the exchange energy
to be used in density functional theory calculations of
two-dimensional systems. This class of approximations has a long and
successful history, but it has not yet been fully investigated for
electrons in two dimensions. We follow the approach originally
proposed by Becke for three-dimensional systems [Int. J. Quantum Chem. {\bf 23}, 1915 (1983),
J. Chem. Phys. {\bf 85}, 7184 (1986)]. The resulting functional depends on two parameters
that are adjusted to a test set of parabolically confined
quantum dots. Our exchange functional is then tested on a variety of
systems with promising results, reducing the error in the
exchange energy by a factor of 4 with respect to the simple local
density approximation.
\end{abstract}

\pacs{31.15.E-, 71.15.Mb}

\maketitle

\section{Introduction}
Present nanoscale electronic devices contain a large variety of
low-dimensional systems in which the many-body problems of interacting
electrons need to be addressed. These systems include, e.g., modulated
semiconductor layers and surfaces, quantum Hall systems, spintronic
devices, and quantum dots~\cite{qd}. In order to describe the
electronic properties of these systems, a practical and accurate way
of computing the energy components is required.  Since the advent of
density-functional theory~\cite{dft1,dft2,dft3} (DFT) much effort went into the
development of approximate functionals for the exchange and
correlation energies. Most of this work focused on three-dimensional
(3D) systems, where considerable advances beyond the commonly used
local density approximation (LDA) were achieved by generalized
gradient approximations (GGAs), orbital functionals, and hybrid
functionals~\cite{functionals}. However, previous studies have shown
that most functionals developed for 3D systems break down when applied to
two-dimensional (2D) systems~\cite{kim,pollack}.

Within the DFT approach, 2D systems are most commonly treated using
the 2D-LDA exchange [see Eq. (13) in Ref.~\onlinecite{rajagopal}], 
which is then combined with the 2D-LDA
correlation parametrized first by Tanatar and Ceperley~\cite{tanatar}
and later, for the complete range of collinear spin polarization, by
Attaccalite and co-workers~\cite{attaccalite}. Despite the relatively
good performance of LDA with respect to, e.g., quantum Monte Carlo
calculations~\cite{henri}, there is a clear lack of accurate 2D
density functionals.

The exact-exchange functional employed within the optimized effective
potential method, which automatically conforms to various
dimensionalities, seems an appealing alternative to the LDA, and it 
has recently
been applied to quantum dots~\cite{nicole}. In that method, however, the
development of approximations for the correlation energies compatible
with exact-exchange energies remains a complicated
problem. 

The derivation of local and semi-local approximations for the
exchange-correlation energy in 2D can be carried out following the
general lines already employed for the 3D case. In this way one can
take advantage of the almost 40 years of experience in that field.
Only very recently, such efforts have been done in 2D by developing
exchange functionals specially tailored for {\em finite} 
2D systems.~\cite{pittalis,ring,ususus,correlation}
In this work we take the most natural step beyond the LDA by
including the density gradients in the functional.
To this end, there are several possible approaches, e.g., the
gradient expansion of the exchange hole~\cite{perdew1}, 
semiclassical expansions from the Dirac or Bloch density 
matrix,~\cite{dft1} and the GGAs 
of Perdew~\cite{perdew2} 
and Becke.~\cite{becke2,becke3,becke88} Here we follow the 
approach introduced by Becke for 3D systems in 
Refs.~~\onlinecite{becke2} and ~\onlinecite{becke3}, 
and derive and apply a GGA for
the exchange energy of 2D electronic systems.
Tests for a diverse set
of 2D quantum dots show excellent performance of the
derived approximation when compared with exact-exchange results.

\section{Derivation of the approximations}

Within the Kohn-Sham approach to spin-DFT~\cite{BarthHedin:72}, the
ground state energies and spin densities $n_{\sigma}(\vr)$ of a system
of $N=N_{\uparrow}+N_{\downarrow}$ interacting electrons are
determined. The total energy, which is minimized to obtain the
ground-state energy, is written as a functional of the spin densities
(in Hartree atomic units)
\begin{multline}
  \label{etot}
  E[n_{\sigma}] = T_\text{s}[n_{\sigma}] + 
  E_{\rm H}[n] + E_\text{xc}[n_{\sigma}] \\
  + \sum_{\sigma=\uparrow,\downarrow}\Id{r} v_{\sigma}(\vr)
  n_{\sigma}(\vr), 
\end{multline}
where $T_\text{s}[n_{\sigma}]$ is the Kohn-Sham
kinetic energy functional, $v_{\sigma}(\vr)$ is the external (local)
spin-dependent scalar potential acting upon the interacting system, $E_{\rm H}[n]$ is
the classical electrostatic or Hartree energy of the total charge
density $n(\vr)=n_{\uparrow}(\vr)+n_{\downarrow}(\vr)$, and
$E_\text{xc}[n_{\sigma}]$ is the exchange-correlation
energy functional. The latter can be further decomposed into the
exchange and correlation parts as $E_\text{xc} = E_\text{x} + E_\text{c}$.

In this work we focus in the exchange-energy functional, that can be expressed as
\begin{equation}
  \label{EX_2}
  E_\text{x} [n_{\sigma}] = - \frac{1}{2} \sum_{\sigma=\uparrow,\downarrow} 
  \Id{r_1}\!\!\Id{r_2} \frac{n_{\sigma}(\vr_1)}{|\vr_1-\vr_2|}
  h_{\text{x},\sigma}(\vr_1,\vr_2),
\end{equation}
where, within the restriction that the noninteracting ground state is
nondegenerate and hence takes the form of a single Slater determinant,
the exchange-hole (or Fermi-hole) function is given
by
\begin{equation}
  \label{xhole}
  h_{\text{x},\sigma}(\vr_1,\vr_2) =
  \frac{|\sum_{k=1}^{N_\sigma}\varphi^*_{k,\sigma}(\vr_1)\varphi_{k,\sigma}(\vr_2)|^2}
  {n_{\sigma}(\vr_1)}.
\end{equation}
The sum in the numerator is the one-body spin-density matrix of the
Slater determinant constructed from the Kohn-Sham orbitals,
$\varphi_{k,\sigma}$. Moreover, integrating this function over $\vr_2$
yields
\begin{equation}
  \label{norm}
  \Id{r_2} h_{\text{x},\sigma}(\vr_1,\vr_2) = 1.
\end{equation}
This exact property reflects the fact that around an electron with
spin $\sigma$ at $\vr_1$, other electrons of the same spin are less
likely to be found. This a consequence of the Pauli principle.  From
Eq.~(\ref{EX_2}) it is clear that to evaluate the exchange energy in
2D, we just need to know the {\em cylindrical} average
w.r.t. $\vs=\vr_2-\vr_1$ of the exchange-hole around $\vr_1$,
\begin{equation}
  \label{avxhole}
  \bar{h}_{\text{x},\sigma}(\vr_1,s) = \frac{1}{2\pi}
  \int_{0}^{2\pi}\!\!{\rm d}\phi_s\:  h_{\text{x},\sigma}(\vr_1,\vr_1+\vs),
\end{equation} 
from which
\be\label{EXCA}
  E_\text{x} [n_{\sigma}] = - \pi \sum_{\sigma=\uparrow,\downarrow}
  \Id{r} n_{\sigma}(\vr)
  \int\!\!{\rm d} s \: \bar{h}_{\text{x},\sigma}(\vr, s),
\ee
where we have renamed $\vr_1$ as $\vr$.  Expressing the exchange hole
by its Taylor expansion, and considering its cylindrical average, one
arrives at the following expression,
\begin{equation}
  \label{taylor2}
  \bar{h}_{\text{x},\sigma}(\vr,s)  =  n_{\sigma}(\vr) + C^{\sigma}_\text{x}(\vr)s^2  + \ldots ,
\end{equation}
where $C^{\sigma}_\text{x}$ is the so-called local curvature of the
exchange hole around the given reference point $\vr$. This function
can be expressed as~\cite{becke2,Dobson:93,pittalis,esa}
\begin{equation}
  \label{C}
  C^{\sigma}_\text{x}(\vr) = \frac{1}{4}\left[ \nabla^2 n_{\sigma}(\vr) -2\tau_\sigma(\vr)
  + \frac{1}{2}\frac{\left| \nabla n_\sigma(\vr) \right|^2}{n_\sigma}
  + 2 \frac{\vj^2_{p,\sigma}(\vr)}{n_\sigma(\vr)} \right],
\end{equation}
where
\begin{equation}
  \tau_\sigma(\vr)=\sum_{k=1}^{N_\sigma} |\nabla\varphi_{k,\sigma}(\vr)|^2
\end{equation}
is (twice) the spin-dependent kinetic-energy density, and
\begin{multline}
  \vj_{p,\sigma}(\vr)=\frac{1}{2i}\sum_{k=1}^{N_\sigma} \Big\{
  \varphi^*_{k,\sigma}(\vr) [\nabla \varphi_{k,\sigma}(\vr)] - \\
  [\nabla \varphi^*_{k,\sigma}(\vr)] \varphi_{k,\sigma}(\vr) \Big\}
\end{multline}
is the spin-dependent paramagnetic current density.  Both
$\tau_\sigma$ and $\vj_{p,\sigma}$ depend explicitly on the Kohn-Sham
orbitals. Thus the expression in Eq.~(\ref{C}) has an implicit
dependence on the spin densities $n_\sigma$.

\subsection{Small density-gradient limit}

When the inhomogeneity of the electron system is small, we may regard the
homogeneous 2D electron gas (2DEG) as a good reference system.  In
this case, we have an exact expression~\cite{gorigiorgi}
\begin{equation}
  \label{2DEG_xhole}
  \bar{h}^{\rm 2DEG}_{\text{x},\sigma}(s)= 
    \frac{ k^{2}_{F,\sigma}}{\pi} \frac{J^2_1(k_{F,\sigma}\; s)}{(k_{F,\sigma}\; s)^2},
\end{equation}
where $k_{F,\sigma}=\sqrt{4\pi n_{\sigma}}$ is the Fermi momentum
(for spin $\sigma$) in 2D, and $J_1$ is the ordinary Bessel function
of the first kind in the first order. Notice that the principal maximum of
Eq.~(\ref{2DEG_xhole}) accounts for $95\%$ of the exchange energy.  We
thus follow the idea introduced by Becke for the 3D case~\cite{becke2}
to modify the principal maximum of $\bar{h}^{\rm 2DEG}_{x,\sigma}$ by a
polynomial factor.  This improves the short-range behavior but leaves
the secondary maximum unchanged.  We may write
\be
  \label{modela}
  \bar{h}_{\text{x},\sigma}(\vr,s) = \left[1+ a_{\sigma}(\vr)s^2 + b_{\sigma}(\vr)s^4 +
  \ldots \right]\bar{h}^{\rm 2DEG}_{\text{x},\sigma}(s),
\ee
for $k_{F,\sigma} s < z$, and
\be
  \label{modelb}
  \bar{h}_{\text{x},\sigma}(\vr,s) = \bar{h}^{\rm 2DEG}_{\text{x},\sigma}(s),
\ee
for $k_{F,\sigma} s > z$, where $z$ is the first zero of $J_1$.

Now we have to find an expression for both $a_{\sigma}$ and
$b_{\sigma}$. Comparing Eq.~(\ref{modela}) with Eq.~(\ref{taylor2}),
making use of the series representation of $J_1$
\be
  J_1(y) = \sum_{k=0}^{\infty} 
  \frac{ \left( -1 \right)^{k} \left( \frac{y}{2} \right)^{2k+1}}{k!\,\Gamma(k+2)},
\ee
and replacing the expression of $\tau_{\sigma}$ in Eq.~(\ref{C}) by
the 2D Thomas-Fermi expression~\cite{Zyl2,Berkane,pittalis}
\be
  \label{te}
  \tau_{\sigma}(\vr) = 2\pi n^2_{\sigma}(\vr)+ \frac{1}{3} \nabla^2 n_{\sigma}(\vr) + 
  \frac{\vj^2_{p,\sigma}(\vr)}{n_\sigma(\vr)}\,,
\ee
we arrive at
\be
  a_{\sigma}(\vr) = \frac{1}{ 4 n_{\sigma}(\vr) } 
  \left[\frac{2}{3} \nabla^2 n_{\sigma}(\vr) 
  + \frac{1}{2} \frac{\left| \nabla n_{\sigma}(\vr) \right|^2}{n_{\sigma}(\vr)}
  \right].
\ee
Next, we can determine $b_{\sigma}$ by applying the normalization
constraint of Eq.~(\ref{norm}), leading to
\be
  b_{\sigma}(\vr)=-4\pi \frac{I(1)}{I(3)} n_{\sigma}(\vr) a_{\sigma}(\vr)
  \,.
\ee
Here the symbol $I(m)$ denotes
\be
  I(m) = \int_{0}^{z}\!\! {\rm d}y\: y^m J^2_1(y)
  \,,
\ee
where $z$ is again the first zero in $J_1$.
The values of the integrals $I(n)$ can be determined numerically.

Making use of Eq.~(\ref{EXCA}), we arrive at 
\be
  E_\text{x}[n_{\sigma}] =  E^{\rm 2DEG}_{\text{x},\sigma}[n_{\sigma}] + 
  E^{\rm SGL}_{\text{x},\sigma}[n_{\sigma}] 
\ee
where
\begin{multline}
  \label{SGL1}
  E^{\rm SGL}_\text{x} [n_{\sigma},\nabla n_{\sigma}] = \\ - \tilde\kappa
  \sum_{\sigma} \Id{r}  n^{-1/2}_{\sigma}(\vr) \left[\frac{2}{3}  \nabla^2 n_{\sigma}(\vr) +
  \frac{1}{2} \frac{\left| \nabla n_{\sigma}(\vr) \right|^2}{n_{\sigma}(\vr)}\right],
\end{multline}
with
\be
  \tilde\kappa = \frac{1}{4^{3/2} \sqrt{\pi}} \left[  \frac{I(0)I(3)-I(1)I(2)}{I(3)} \right],
\ee
where SGL refers to the small-gradient limit.  Using Green's first
identity when integrating the first term, we find
\be\label{SGL}
  E^{\rm SGL}_{\text{x},\sigma} 
  = -\kappa_{\rm SGL}
  \Id{r} n^{3/2}_{\sigma}(\vr) x^2_\sigma(\vr)\,,
\ee
with $\kappa_{\rm SGL} = 5\tilde\kappa/6$ and $x_\sigma(\vr)=\left| \nabla n_{\sigma}(\vr) \right|
/ n^{3/2}_{\sigma}(\vr)$ is the usual dimensionless parameter for exchange.

\subsection{Large density-gradient limit}

Using Eqs.~(\ref{C}) and (\ref{te}), we can rewrite
Eq.~(\ref{taylor2}) as
\begin{multline}
  \bar{h}_{\text{x},\sigma}(\vr,s)  \approx  n_{\sigma}(\vr) \\  
  +
  \frac{1}{4}\left[ \nabla^2 n_{\sigma}(\vr) -4\pi n^2_{\sigma}(\vr)
  + \frac{1}{2}\frac{\left| \nabla n_\sigma(\vr) \right|^2}{n_\sigma (\vr)} \right] s^2 .
\end{multline}
Here we follow the reasoning of Becke~\cite{becke3} (applied in 3D),
and assume that the term in the density gradient dominates over the
other terms. Thus, we obtain
\be
  \label{lg1}
  \bar{h}_{\text{x},\sigma}(\vr,s)  \approx  
  \frac{1}{8}\frac{\left|\nabla n_\sigma(\vr) \right|^2}{n_\sigma(\vr)}  s^2.
\ee
We note that expression (\ref{lg1}) is valid only for small $s$. 
Otherwise, following again the argument of Becke, we propose 
\be
  \label{lg2}
  \bar{h}_{\text{x},\sigma}(\vr,s) =
  \left[\frac{1}{8} \frac{\left|\nabla n_\sigma(\vr) \right|^2}{n_\sigma(\vr)} s^2 \right]
  F(\alpha_{\sigma}(\vr)\, s),
\ee
where
\be
  \label{F}
  F(y) = e^{-y^2}.
\ee
The form of $F$ in Eq.~(\ref{F}) corresponds to a Gaussian
approximation for the exchange-hole. This would be exact in the case
of a single electron in a harmonic confinement potential. However,
another choice for $F$ may be considered; so far it allows to
reproduce the correct short-range behavior given by Eq.~(\ref{lg1}),
and decay in a way leading to finite exchange energies.

In Eq.~(\ref{lg2}), the parameter $\alpha_\sigma$ can be determined by
enforcing the normalization condition of Eq.~(\ref{norm}). This leads
to
\be
  \alpha^4_{\sigma}(\vr) = 
  \frac{\pi~G(3)}{4} \frac{\left|\nabla n_\sigma(\vr) \right|^2}{n_\sigma(\vr)}
  \,,
\ee
where 
\be
  G(m) = \int_{0}^{\infty}\!\!{\rm d} y\: y^m e^{-y^2}.
\ee

Finally, making use of Eq.~(\ref{EXCA}), we arrive at 
\be
  \label{LGL}
  E^{\rm LGL}_{x,\sigma}[n_{\sigma},\nabla n_{\sigma}] 
  =-\kappa_{\rm LGL} \Id{r} n^{3/2}_{\sigma}(\vr) x^{1/2}_\sigma(\vr),
\ee
where
\be
\kappa_{\rm LGL} = \frac{\pi^{1/4}}{2^{3/2}}G(2)G^{-3/4}(3).
\ee
Here LGL refers to the large-gradient limit.

\section{Exchange-energy functional}

\begin{figure}
\setlength{\unitlength}{0.95\columnwidth}
\begin{center}
  \includegraphics[width=\unitlength]{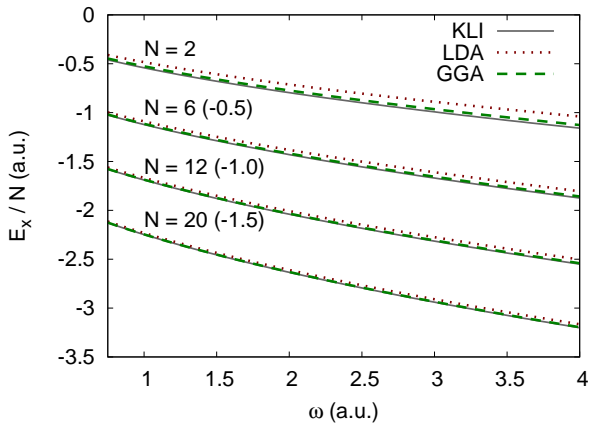}
  \vspace{-1cm}
\end{center}
\caption{\label{fig:vs_w}
(color online) 
Exchange energy per electron for a series of parabolically confined
quantum dots with $N$ electrons and confinement strength $\omega$. For
clarity, the line for $N=6$ was shifted down by 0.5\,a.u., the line
for $N=12$ by 1\,a.u., and the line for $N=20$ by 1.5\,a.u. The
lines are: exact exchange (solid gray), exchange-only LDA (dotted
red), and this work (dashed green).}
\end{figure}

We can combine Eqs.~(\ref{SGL}) and (\ref{LGL}) in an expression that
interpolates the exchange energy of an inhomogeneous electron gas from
the small density-gradient limit to the large one. A possible
expression is of the form
\begin{multline}
  E^{\rm GGA}_\text{x}[n_{\sigma},\nabla n_{\sigma}] =
  E^{\rm LDA}_\text{x}[n_{\sigma}] \\ 
  -\beta\sum_{\sigma=\uparrow,\downarrow}
  \Id{r} n^{3/2}_{\sigma}(\vr) \frac{x^2_{\sigma}(\vr)}
  {\left[ 1 + \gamma x^2_{\sigma}(\vr)\right]^{3/4}},
\end{multline}
or, writing this expression explicitly
in terms of the density and its gradients,
\begin{multline}
 E^{\rm GGA}_\text{x}[n_{\sigma},\nabla n_{\sigma}] =
E^{\rm LDA}_\text{x}[n_{\sigma}] \\
-\beta\sum_{\sigma=\uparrow,\downarrow}
\Id{r} \frac{|\nabla n_\sigma(\vr)|^2}{n^{3/2}_{\sigma}(\vr)\left[1+\gamma\frac{|\nabla n_\sigma(\vr)|^2}{n^3_{\sigma}(\vr)}\right]^{3/4}}\,,
\label{functional}
\end{multline}
where $\beta$ and $\gamma$ can be trivially written in terms of
$\kappa_{\rm LGL}$ and $\kappa_{\rm SGL}$. At last, to improve the
flexibility of the above approximate functional, we replace both
$\beta$ and $\gamma$ with two parameters to be determined by fitting
the exchange energies of an ensemble of physically relevant 2D
systems.

We have chosen to fit our parameters to a set of four 
parabolic two-electron quantum dots~\cite{qd} with confinement stengths
$\omega = 1,\: 1/4,\: 1/16$, and 1/36\,a.u., respectively. 
These are very well studied systems~\cite{rontani} that span a wide 
range of the density parameter, $0.9 \lesssim r_s \lesssim 5.7$, where
$r_s=N^{-1/6}\,\omega^{-2/3}$. 
For the reference exchange energies we
performed self-consistent exact-exchange calculations with the
code {\tt octopus}\cite{octopus} using the
Krieger-Li-Iafrate~\cite{KLI} (KLI) approximation, which is a very
accurate approximation in the static case~\cite{kli_oep}. We obtained
$\beta=0.003317$ and $\gamma=0.008323$, which are close to
the parameters found by Becke for the 3D case by fitting a series of
rare-gas atoms, $\beta^\text{3D}=0.00375$ and
$\gamma^\text{3D}=0.007$ (Ref.~~\onlinecite{becke3}).

\begin{table}[t]
  \caption{\label{table_qd} Exchange energies (in
  atomic units) for parabolic quantum dots. The last row contains the
  mean percentage error, $\Delta$. The first 
  four lines represent the systems that were used in the fitting of 
  $\beta$ and $\gamma$. The columns are: exact exchange 
  (EXX), local density approximation (LDA), and the 
  generalized gradient approximation presented in this work (GGA).
}
  \begin{tabular}{c c c c c}
  \hline
  \hline
  $N$ & $\omega$ & $-E^{\rm EXX}_{\rm x}$ &  $-E_{\rm x}^{\rm LDA}$ & $-E_{\rm x}^{\rm GGA}$   \\
  \hline
  2  & 1           & 1.083  & 0.9672 & 1.051  \\
  2  & 1/4         & 0.4850 & 0.4312 & 0.4704 \\
  2  & 1/16        & 0.2073 & 0.1843 & 0.2023 \\
  2  & 1/36        & 0.1239 & 0.1108 & 0.1276 \\
  6  & 0.42168     & 2.229  & 2.110  & 2.206  \\
  6  & $1/1.89^2$  & 1.735  & 1.642  & 1.719  \\
  6  & 1/4         & 1.618  & 1.531  & 1.603  \\
  12 & $1/1.89^2$  & 3.791  & 3.668  & 3.777  \\
  \hline
  \multicolumn{3}{l}{$\Delta$} & 7.9\% & 1.8\% \\
  \hline
  \hline
  \end{tabular}
\end{table}

\begin{table}
  \caption{\label{table_qd_recta} Exchange energies (in
  atomic units) for square (${\rm area}=\pi\times\pi$) quantum dots. The meaning of 
  the columns is identical to Table~\ref{table_qd}.
}
  \begin{tabular}{c c c c}
  \hline
  \hline
  $N$ & $-E^{\rm EXX}_{x}$ & $-E_{\rm x}^{\rm LDA}$ & $-E_{\rm x}^{\rm GGA}$ \\
\hline
   2 & 1.417 & 1.288 & 1.383 \\
   6 & 6.147 & 5.902 & 6.180 \\
   8 & 9.509 & 9.017 & 9.434  \\
  12 & 16.24 & 15.91 & 16.46 \\
  16 & 25.23 & 24.35 & 25.15 \\
  \hline
  \multicolumn{2}{l}{$\Delta$} & 4.8\%  & 1.1\% \\
  \hline
  \hline
  \end{tabular}
\end{table}

Our functional was then applied to parabolically confined quantum dots
by varying the number of electrons $N$ in the dot and the confinement
strength $\omega$. From the fully self-consistent DFT 
results summarized in Fig.~\ref{fig:vs_w}
it is clear that our functional outperforms the LDA for the whole
range of $N$ and $\omega$. A more quantitative picture can be obtained
from Tables~\ref{table_qd} and \ref{table_qd_recta} where we present
numerical values for the exchange energy obtained with different
approximations, for both parabolically confined and hard-wall 
square~\cite{square} quantum dots, respectively.
It is clear that our functional yields errors that are 
smaller by at least a factor of 4 than the errors 
of the simple LDA. 

We tested the performance of our GGA functional also in the
large-$N$ limit. Here we used the exact exchange energies known for
parabolic closed-shell quantum dots of noninteracting electrons.~\cite{zyl}
Using a {\em fixed}, analytic electron density as an input in
Eq.~(\ref{functional}), 
we found relative percentage errors of $0.7\,\%$ and $0.5\,\%$ for
$N=110$ and $420$, respectively, whereas the corresponding errors of
the LDA are $0.5\,\%$ and $0.2\,\%$. Even if the plain LDA is slightly
more accurate in this
case, we find the result very satisfactory in view of the fitting
of $\beta$ and $\gamma$ according to {\em two-electron} data as
explained above. We point out that in terms of potential 
applications for our GGA functional, the main interest is in
relatively small systems (up to a few dozen of electrons).
However, the GGA described here -- possibly with simple 
modifications -- enables precise many-electron calculations also in 
large-scale structures such as quantum-dot arrays~\cite{zaremba} or 
quantum-Hall devices.~\cite{siddiki}

\section{Conclusions}

We proposed a generalized gradient approximation for the exchange energy 
of two-dimensional systems, following the same lines developed for three-dimensional 
systems proposed by Becke~\cite{becke2,becke3}. Analyzing the 
small- and large-density gradient limits, we arrived at 
the expression of a functional which depends on two
parameters, that are fitted to a test set composed of four parabolically confined quantum dots 
containing two electrons. Further calculations, both for parabolically
confined and square quantum dots, show that our
approximations yields errors that are at least a factor of 4 better
than the local density approximation. 

We believe that this is the necessary step in the construction
of reliable generalized gradient approximations for two-dimensional systems.
The next steps will necessarily involve the construction of correlation functionals
(beyond the local density approximation~\cite{attaccalite} or the 
forms proposed in Refs.~~\onlinecite{ususus} and ~\onlinecite{correlation}),
improvements to the exchange functional
presented in this work, and further extensive tests, and analysis, to assess the
quality of the same functionals. In this path we expect that the
experience in the development of functionals in
three dimensions will be a very useful guide, but we can also expect
that experience gained in describing exchange and correlation in these low-dimensional
systems can be again transposed to bring new insights and ideas into
the three-dimensional world.

\begin{acknowledgments}
This work was supported by the EU's Sixth Framework Programme through
the Nanoquanta Network of Excellence (NMP4-CT-2004-500198), Deutsche
Forschungsgemeinschaft, the Academy of Finland, 
and the Funda{\c c}{\~a}o para a Ci{\^e}ncia e Tecnologia 
through Project No. SFRH/BD/38340/2007. M.A.L.M.
acknowledges partial support by the Portuguese FCT through Project No.
PTDC/FIS/73578/2006. 
\end{acknowledgments}

\end{document}